\documentclass[10pt]{article}
\usepackage{graphicx,floatflt,amssymb,epsfig,epsf} 
\def\be {\begin{equation}}
\def\ee {\end{equation}}
\def\bea {\begin{eqnarray}}
\def\eea {\end{eqnarray}}

\def\ztwo {{\cal Z}_2}

\def\pir {\pi R}
\def\stwo {\sqrt{2}}
\def\spir {\sqrt{\pir}}
\def\mhsqbar {{\bar{m}_h}^2}

\def\opcit(#1){ {\em op. cit.}, #1}

\def\issue(#1,#2,#3){#1, #2 (#3)} 

\def\APP(#1,#2,#3){Acta Phys.\ Polon.\ \issue(#1,#2,#3)}
\def\ARNPS(#1,#2,#3){Ann.\ Rev.\ Nucl.\ Part.\ Sci.\ \issue(#1,#2,#3)}
\def\CPC(#1,#2,#3){Comp.\ Phys.\ Comm.\ \issue(#1,#2,#3)}
\def\CIP(#1,#2,#3){Comput.\ Phys.\ \issue(#1,#2,#3)}
\def\EPJC(#1,#2,#3){Eur.\ Phys.\ J.\ C\ \issue(#1,#2,#3)}
\def\EPJD(#1,#2,#3){Eur.\ Phys.\ J. Direct\ C\ \issue(#1,#2,#3)}
\def\IEEETNS(#1,#2,#3){IEEE Trans.\ Nucl.\ Sci.\ \issue(#1,#2,#3)}
\def\IJMP(#1,#2,#3){Int.\ J.\ Mod.\ Phys. \issue(#1,#2,#3)}
\def\JHEP(#1,#2,#3){J.\ High Energy Physics \issue(#1,#2,#3)}
\def\JPG(#1,#2,#3){J.\ Phys.\ G \issue(#1,#2,#3)}
\def\MPL(#1,#2,#3){Mod.\ Phys.\ Lett.\ \issue(#1,#2,#3)}
\def\NP(#1,#2,#3){Nucl.\ Phys.\ \issue(#1,#2,#3)}
\def\NIM(#1,#2,#3){Nucl.\ Instrum.\ Meth.\ \issue(#1,#2,#3)}
\def\PL(#1,#2,#3){Phys.\ Lett.\ \issue(#1,#2,#3)}
\def\PRD(#1,#2,#3){Phys.\ Rev.\ D \issue(#1,#2,#3)}
\def\PRL(#1,#2,#3){Phys.\ Rev.\ Lett.\ \issue(#1,#2,#3)}
\def\SJNP(#1,#2,#3){Sov.\ J. Nucl.\ Phys.\ \issue(#1,#2,#3)}
\def\ZPC(#1,#2,#3){Zeit.\ Phys.\ C \issue(#1,#2,#3)}

\begin{document} 
%
%


\begin{flushright} 
CU-PHYSICS/06-2006\\
\end{flushright} 
 
\vskip 30pt 
 
\begin{center} 
{\Large \bf The Excited Scalars of the Universal Extra Dimension Model}\\
\vspace*{1cm} 
\renewcommand{\thefootnote}{\fnsymbol{footnote}} 
{\large {\sf Biplob Bhattacherjee}${}^1$ and {\sf Anirban Kundu}${}^{1,2}$ } \\ 
\vspace{10pt} 
{\small 
   {${}^1$ \em Department of Physics, University of Calcutta, 92 A.P.C. 
        Road, Kolkata 700009, India}\\
   {${}^2$ \em Abdus Salam International Centre for Theoretical Physics,
               Trieste, Italy}}
 
\normalsize 
\end{center} 
 
\begin{abstract} 
In the minimal Universal Extra Dimension (mUED) model, there are four physical
scalar particles at the $n=1$ level, two charged and two neutral. Due to
the almost degenerate nature of the spectrum, the detection of these scalars
is a major challenge, perhaps the greatest experimental challenge if UED-type
new physics is observed at the Large Hadron Collider (LHC). We explore the 
possibility of detecting these particles at the International Linear Collider 
(ILC), and emphasise the need of having an excellent soft $\tau$ detection 
efficiency.
\end{abstract}

\vskip 5pt \noindent 
\texttt{PACS numbers:~ 11.25.Mj, 14.80.Cp, 13.66.Fg} \\ 
\texttt{Keywords:~~Universal Extra Dimension, Electron-Positron Collider,
Excited Higgs Bosons}

\renewcommand{\thesection}{\Roman{section}} 
\setcounter{footnote}{0} 
\renewcommand{\thefootnote}{\arabic{footnote}} 

\section{Introduction}

The possibility of a compactified extra dimension was first discussed
by Nordstr\"om, Kaluza and Klein \cite{kaluza} \footnote{It is
unfortunate that Nordstr\"om hardly gets his due credit.}. 
Such extra dimension (ED) models
were later revived by the necessity of a consistent formulation of
string theories \cite{julia}. Of course, even without string theories,
the world may have one or more compactified dimensions. 
There are a number of such ED models, and
they differ mainly in two ways: first, the number of EDs, the
geometry of space-time, and
the compactification manifold, and second, which particles can go into
the extra dimensions (hereafter called bulk) and which cannot. 

In the so-called Universal Extra Dimension (UED) model
proposed by Appelquist, Cheng, and Dobrescu \cite{acd}, 
all SM particles can go into the bulk. 
In the simplest UED scenario, there is only one
extra dimension, denoted by $y$, compactified on a circle ($S_1$) of
radius $R$. The model predictions
remain essentially unchanged if there are more extra dimensions, 
with a hierarchical radii of compactification. We will focus on 
the simplest scenario only. 

To get chiral fermions at low-energy, one must impose a further $\ztwo$
symmetry ($y\leftrightarrow -y$), so that finally we have an $S_1/\ztwo$
orbifold. As is well-known, a higher dimension theory is nonrenormalizable
and should be treated in the spirit of an effective theory  valid upto
a scale $\Lambda > R^{-1}$. All fields have five space-time components;
when brought down to four dimensions, for each low-mass (zero-mode) Standard
Model (SM) particle of mass $m_0$, 
we get an associated Kaluza-Klein (KK) tower, the
$n$-th level (this $n$ is the KK number of the particle)
of which has a mass given by
\be
m_n^2 = m_0^2 + \frac{n^2}{R^2}.
  \label{kktree}
\ee
This is a tree-level relationship and gets modified once we take into
account the radiative corrections.

Another important feature of the UED scenario is the conservation of the
KK number. This is simply a reflection of the fact that all particles can go
into the fifth dimension and so the momentum along the fifth dimension
must be conserved. Also, this means that the lowest-mass $n=1$ particle,
which turns out to be the $n=1$ photon, is absolutely stable. Such a
lightest KK particle (LKP), just like the lightest supersymmetric particle
(LSP), is an excellent candidate for dark matter \cite{servant,mazumder,byrne,
kk2dark,kongmatchev}. In fact, existence of such a dark matter 
candidate is a general 
property of any theory with such a $\ztwo$ symmetry. To us, this seems
to be the greatest theoretical motivation for UED-type models. 

Radiative corrections to the masses of the KK particles have been computed
in \cite{georgi,cheng1,pk}. These papers, in particular \cite{cheng1}, show
that the almost mass-degenerate spectrum for any KK level, resulting from
eq.\ (\ref{kktree}), splits up due to such correction terms. There are two
types of correction, arising from the fact the fifth dimension is inherently
different from the four large dimensions (a breaking of the Lorentz invariance).
The first one, which results just from the compactification
of the extra dimension, is in general small (zero for fermions) and is
constant for all $n$ levels. This we will call the bulk correction, which arises
when the loops can sense the compactification ({\em e.g.}, loops with
nonzero winding number around $y$). The second
one, which we will call boundary correction, is comparatively large (goes
as $\ln\Lambda^2$ and hence, in principle, can be divergent), and plays
the major role in  determining the exact spectrum and possible decay modes.
The boundary correction terms are related with the interactions present only
at the fixed points $y=0$ and $y=\pi R$. If the interaction is symmetric 
under the exchange of these fixed points (this is another $\ztwo$ symmetry,
but not the $\ztwo$ of $y\leftrightarrow -y$), the conservation of KK
number breaks down to the conservation of KK parity, defined as $(-1)^n$.
Thus, LKP is still stable, but one can in principle have KK-number violating
vertices in the theory, where the violation is by an even amount.
For example, it is possible to produce an $n=2$ state
from two $n=0$ states.

The role of linear colliders in precision study of TeV-scale extra
dimension models has been already emphasized in the literature
\cite{antoniadis}.
The low-energy phenomenology has been discussed in \cite{acd,
agashe,chk,buras,papavas,oliver,ewued,recentpheno}, 
and the high-energy collider signatures
in \cite{rizzo,nandi1,nandi2,bdkr,asesh,cheng2,riemann,bk1,rai1}. 
The limit on $1/R$ from
precision data is about 250-300 GeV, while the limit estimated from dark
matter search \cite{servant} is about a factor of two higher. 
The loop corrections are quite insensitive
to the precise values of the radiative corrections. With the proposed reach
of the International Linear Collider (ILC)
 in mind, we will be interested in the range 300 GeV $<R^{-1}<$ 500 GeV. 

It was first pointed out in \cite{cheng2} that the signals of the UED models
may mimic those of R-parity conserving supersymmetry (SUSY), mainly because
of the $\ztwo$ symmetry of both the theories that predicts a missing energy
signal once a sparticle or a KK excitation is produced. The decay patterns
of excited fermions and sfermions are also similar for a significant part of
the SUSY parameter space. One may hope to discriminate these two models 
by measuring the spin of the excited particles \cite{barr,webber,0509246},
or by producing the $n=2$ gauge bosons as $s$-channel resonances \cite{bk1}.
The standard verdict is that the excited states are to be first seen at the 
Large Hadron Collider (LHC), and a precision study is to be done at 
the ILC. 

One important point that has not been addressed is the production and 
subsequent decay and detection of the $n=1$ scalars. This is perhaps the
most challenging part of a complete determination of the KK spectrum.
There are four such excited scalars at the $n=1$ level, two charged and
two neutral. Not only they are almost mass degenerate, it appears from a 
study of the spectrum that their only possible decay channel is to the
$n=1$ $\tau$ leptons. Ultimately, what we will be left with are 2-4 soft
$\tau$ leptons plus a huge amount of missing energy. These $\tau$s may be
so soft as to avoid detection. Thus, if we are unlucky, {\em the whole 
excited scalar sector} may be invisible! This is to be compared with the 
SUSY models, where one is almost certain to detect some of the Higgs bosons
if they are produced. A detailed discussion of this follows in Sections 2
and 3.

Even if the soft $\tau$'s are visible (and for most of the parameter space
they are), the background is still severe. The most important background
comes from another UED process, the $s$-channel production of excited
$\tau$-lepton pair, which subsequently decays to a pair of $n=0$ soft $\tau$'s
and two LKPs. The signal is indeed a fraction of the background, so we
need some luck to detect the signal. There are other backgrounds, but 
for most of the parameter space they can be brought down by suitable cuts. 

The minimal UED model is completely specified by three input parameters:
$1/R$, inverse of the compactification radius, $\Lambda$, the effective
cutoff of the theory, and $\mhsqbar$, the boundary mass squared (BMS)
term for the excited scalars. 
The last parameter can be determined if we can have a 
precision study of the excited scalar sector. This appears very tough, if not
outright impossible, even at the ILC. However, we will derive a theoretical
bound on $\mhsqbar$ from the fact that the $n=1$ charged Higgs boson must 
not be the LKP. This occurs for large negative values of $\mhsqbar$, of the
order of $-(10^3$-$10^4)$ GeV$^2$. 

The paper is arranged as follows. In Section 2 we discuss, briefly, the
salient features of the excited Higgs sector of the minimal UED model.
In Section 3, we discuss the important two- and three-body 
production processes involving one or more excited scalars, and show their
subsequent decay chains. We also estimate their production cross-section
as a function of $1/R$, $\Lambda$ and $\mhsqbar$. This gives us an idea
of the parameter space where such excited bosons can be detected and where 
they just cannot be. The possible backgrounds are also shown. In Section 4,
we find the theoretical lower bound on $\mhsqbar$ from the non-LKP 
condition. We summarize and conclude in Section 5.

Let us mention here that though
the main focus is on the ILC, an identical study may be performed for
CLIC, the proposed multi-TeV $e^+e^-$ machine, with an optimised $\sqrt{s}
=3$ TeV (and may be upgraded to 5 TeV), and luminosity of $10^{35}$ cm$^{-2}$
s$^{-1}$. The electron beam at CLIC may be polarised upto 80\%,
and the positron beam upto 60-80\%, from Compton scattering off a high power
laser beam \cite{clic}. Clearly, the reach of CLIC will be much higher.

\section{Scalar Sector of UED}

Every field in 5 dimension ($x^\mu,y$) can be decomposed in terms of even and
odd components:
\bea
\phi_+(x^\mu,y) &=& \frac{1}{\spir} \phi_+^{(0)}(x^\mu) + \frac{\stwo}{\spir}
\sum_{n=1}^\infty \cos\frac{ny}{R} \phi_+^{(n)} (x^\mu),\nonumber\\
\phi_-(x^\mu,y) &=& \frac{\stwo}{\spir}
\sum_{n=1}^\infty \sin\frac{ny}{R} \phi_-^{(n)} (x^\mu).\eea
At orbifold fixed points $y=0$ and $y=\pi R$, the odd fields are projected out. 
Also, any vertex must contain an even number of odd fields to survive under
integration over $y$. For the fermions, left-chiral doublets and right-chiral
singlets are even, while the opposite combinations are odd. It is clear that
the Higgs fields must be even to maintain the proper Yukawa couplings. 

In the $n=1$ level, there are four scalar fields: $H^0$, $\chi^0_1$, 
and $\chi^\pm_1$ \footnote{We denote the $n=0$ Higgs boson by $h$.}.
The superscript refers to the charge whereas the subscript refers to the KK
number. The last three fields are just the excitations of the $n=0$ Goldstone
bosons. There are three more color-neutral scalars, coming from the fifth
components of the gauge fields $W^\pm$ and $Z$. They are $\ztwo$-odd 
fields, and can occur
first only at the $n=1$ level. They should mix with the Goldstone excitations,
and a combination of them will act as the Goldstone boson of the $n=1$ level.
For the neutral fields, this combination, for the $n$-th level, is given by
\be
G^0_n=\frac{1}{m_{Z_n}}\left[ m_Z \chi^0_n-\frac{n}{R}Z^5_n\right],
\ee
and a similar expression for the charged Goldstones. The orthogonal combinations
will remain as the physical scalar fields, and we will call them $A^0$, and $H^\pm$.
It is clear that if $1/R \gg m_{W,Z}$, the Goldstones are esentially the fifth
components of the gauge bosons, whereas the physical scalars are the excitations
of the $n=0$ Goldstones (and the $n=0$ Higgs boson).   
 
In the absence of radiative corrections, the tree-level masses of the excited
scalars are given by
\be
m_{H_n^0,A_n^0,H_n^\pm}^2= m_n^2+m_{h,Z,W^\pm}^2,
   \label{higgstree}
\ee
but this is modified by radiative corrections, whose effect is simply to add
a universal term $\delta m_H^2$ to the right-hand side of eq.\ (\ref{higgstree})
\cite{cheng1}. The radiative correction is given by
\be
\delta m_H^2 = m_n^2\left[\frac{3}{2}g_2^2+\frac{3}{4}{g'}^2 -\lambda\right]
\frac{1}{16\pi^2}\ln\frac{\Lambda^2}{\mu^2} + \mhsqbar,
   \label{radcorhiggs}
\ee
where $g'$ and $g_2$ are the $U(1)_Y$ and $SU(2)_L$ gauge couplings 
respectively, and $\lambda$ is the self-coupling of the Higgs boson, 
given by $m_h^2=\lambda v^2$ where $v=246$ GeV. $\Lambda$ is the 
effective cutoff scale and $\mu$ is the regularization scale, which 
may be taken for our case to be $1/R$. The term
$\mhsqbar$ is arbitrary; this is the BMS term for the excited scalars, and is
not {\em a priori} calculable. Along with $1/R$ and $\Lambda$, $\mhsqbar$ forms
the complete set of input parameters to specify the minimal 
UED model (of course, one needs to know the $n=0$ Higgs boson mass, $m_h$).
Note that the hierarchy $m_{H^0} > m_{A^0} > m_{H^\pm}$ is fixed.

The necessary Feynman rules are easy to derive; they come from the kinetic term
$(D_M\Phi)^\dag(D^M\Phi)$ of the Higgs field, where $M$ runs over all the five 
dimensions. After compactification, this generates the couplings of the $n=0$ Higgs
boson with a pair of excited gauge bosons, with a pair of the fifth component of
the gauge bosons, and, the couplings of the excited Higgs bosons to $n=0$ and 
excited gauge bosons. As we have said earlier, no vertex can have an odd number of
$\ztwo$-odd fields, and Higgs excitations are even, so there is no need to
take the fifth components of the gauge fields into account (this can be ensured
further by drawing the Feynman diagrams and noting that $n=0$ fields are always
$\ztwo$-even).

\section{Collider Prospects}

Before we go on to the production processes, let us see what these 
excited scalars should decay into. Unless we go to some bizarre corner 
of the parameter space with inordinately large value of $\mhsqbar$, 
these scalars are always lighter than $n=1$ quarks, and so cannot 
decay into them. Similarly, they are also lighter than the excited 
$W^\pm$ and $Z$. The only $n=1$ states that may be below these bosons
are the excited leptons (including neutrinos) and $\gamma_1$, the LKP. 
So, almost all the time, they will decay to the excited $\tau$ lepton 
(remember that KK-number must be conserved). 

For large enough values of $\mhsqbar$, all the $n=1$ scalars can be 
more massive (albeit very slightly) than $W^\pm_1$. However, we have checked
that the splitting is almost never sufficient to have a $n=0$ boson at the final
state. A possibility is a three-body decay to a $n=1$ gauge boson and a
pair of $n=0$ leptons. With no guarantee of having a soft $\tau$ in the final
state, such signals are even more difficult to trace than those we will
discuss here.

There are two excited $\tau$ leptons, which we will call $\tau_1$ and $\tau_2$
(not to be confused with the $n=2$ excitation of $\tau$; we use this notation
since we will never talk about $n\geq 2$ excitations in this paper). 
Both of them are vectorial in nature, but for $\tau_1$, the $SU(2)$ singlet, the
right-chiral state is even while the left-chiral state is odd under $\ztwo$.
For $\tau_2$ it is just the opposite \footnote{The physical states are not
exactly $\tau_1$ and $\tau_2$, but very close to them. This is due to the off-diagonal
term in the $n=1$ $\tau$ mass matrix, which is $m_\tau$, and can be safely 
neglected compared with the diagonal term $1/R$. We do not make such 
hair-splitting distinctions. The analogy with Supersymmetry does not hold, 
since there is no $\tan\beta$ term to enhance the off-diagonal entries.}. 
However, all scalars are $\ztwo$-even, and
they must decay to $\tau_1$ or $\tau_2$ accompanied by a $n=0$ $\tau$ (or $\nu_\tau$),
so $\ztwo$-odd $\tau$s are not produced from $n=1$ Higgs decays.  

Thus, a neutral excited scalar will decay to an excited $\tau^\pm$ and an $n=0$
$\tau^\mp$. The excited $\tau$s will further decay to the ordinary $\tau$
and the LKP. The final result of the cascade is two unlike sign $\tau$ leptons
and an LKP. Since the LKP will carry most of the energy (the typical mass
difference between LKP and excited scalars is $\sim 10$ GeV), both these
$\tau$s will be soft. 

What about the chirality of these $\tau$s? (Since they are soft, helicity 
and chirality can be different.) The coupling of the scalar with the fermions
being Yukawa in nature, both right-chiral $\tau_1$ and left-chiral $\tau_2$
can be produced, accompanied with a $n=0$ $\tau$ of opposite chirality.
It can be seen from Table \ref{masses} 
that $\tau_2$ is closer to $H^0$ and $A^0$.
So, from kinematic considerations, more $\tau_1$s will be produced. It can
so happen that the $\tau_2$ window is completely closed. 

\begin{table}[htbp]
\begin{center}
\begin{tabular}{||c|c|c||c|c|c|c|c|c||}
\hline
$1/R$ & $\Lambda R$ & $\mhsqbar$ & $H^0$ & $A^0$ & $H^\pm$ & LKP & $\tau_2$ &
$\tau_1$ \\
\hline
300 & 20 & 0        & 325.70 & 316.22 & 313.22 & 301.54 & 308.95 & 303.27\\
300 & 20 & $-$5000  & 317.93 & 308.22 & 305.13 & 301.54 & 308.95 & 303.27\\
300 & 20 & 5000     & 333.29 & 324.03 & 321.10 & 301.54 & 308.95 & 303.27\\
300 & 50 & 0        & 326.49 & 317.04 & 314.04 & 301.71 & 311.68 & 304.28\\
300 & 50 & $-$5000  & 318.74 & 309.05 & 305.97 & 301.71 & 311.68 & 304.28\\
300 & 50 &    5000  & 334.06 & 324.83 & 321.90 & 301.71 & 311.68 & 304.28\\
450 & 20 & 0        & 469.77 & 463.25 & 461.20 & 451.07 & 463.42 & 454.91\\
450 & 20 & $-$5000  & 464.42 & 457.82 & 455.75 & 451.07 & 463.42 & 454.91\\
450 & 20 &    5000  & 475.06 & 468.62 & 466.59 & 451.07 & 463.42 & 454.91\\
450 & 50 & 0        & 471.00 & 464.50 & 462.46 & 451.11 & 467.52 & 456.41\\
450 & 50 & $-$5000  & 465.66 & 459.08 & 457.02 & 451.11 & 467.52 & 456.41\\
450 & 50 &    5000  & 476.28 & 469.85 & 467.83 & 451.11 & 467.52 & 456.41\\
\hline
\end{tabular}
    \label{masses}
\caption{The masses of $n=1$ scalars. Also shown are the mass of the LKP,
$\gamma_1$, and the masses of the two $n=1$ $\tau$'s (all in GeV).
The $n=1$ neutrino is degenerate with $\tau_2$. $m_h=120$ GeV.}
\end{center}
\end{table}

For large negative values of $\mhsqbar$ ($\sim -10^4$), $A^0$, and even
$H^0$, can go down below the $\tau_1$ threshold. In that case they will
be comparatively long-lived, decaying into the LKP plus a 
fermion-antifermion pair at $n=0$, or simply to a photon and LKP, through
a triangle diagram. 

The charged boson, $H^+$, will decay into a $\tau^+$ and a $\nu_\tau$. The 
neutrino, whether $n=0$ or $n=1$, is left-chiral, so the $\tau$ must be
right-chiral in nature. The $n=1$ neutrino is unobservable (it decays to
a $n=0$ neutrino and the LKP; note that LKP can directly couple to the
neutrino since it is almost an excited $B$, the hypercharge gauge boson),
and can be called a virtual LKP. 
Other decay channels are kinematically forbidden (like $H^+\to W^+ \gamma_1$)
or highly suppressed (like $H^+\to \mu_1^+ \nu_\mu$), so we will not consider
them any further.

For large negative values of $\mhsqbar$, the $\tau$ window may be closed. In
that case the only possible decay option is $H^+ \to {\rm LKP} + f + \bar{f'}$
through a virtual $n=0$ $W^+$. Since LKP has a very small $W_1$-admixture,
this decay is bound to have a long lifetime, and may even leave a visible
charged track.

In summary, this means that (unless we entertain the possibility of large
negative $\mhsqbar$) from pair production of $H^+ H^-$ one gets a 
pair of unlike sign soft $\tau$s plus missing energy. For the neutral pair
$H^0 A^0$ one gets four soft $\tau$s plus
missing energy. The problem is that some of these $\tau$s (particularly
for the $4\tau$ final state) can be so soft as to miss detection. 

In fig.\ 1 we show the production cross-section for the charged Higgs 
pair as a function of $1/R$ for $\Lambda R=20$ (the values are not much
sensitive on the cutoff). The numerical cumputations were done with the
CalcHEP package \cite{calchep}, augmented by the implementation of UED.
Figures 2-4 show the cross-section for other
rare processes, most of them being hopelessly small, even at ILC (because
of backgrounds that are hard to remove, more on this later). 
$H^0$ is mostly produced through
the fusion of two vector bosons, one of $n=0$ and the other of $n=1$, 
associated with two neutrinos or two electrons. As is expected, the
cross-sections for those
processes that occur via $s$-channel exchanges (like the Bjorken 
process, $e^+e^-\to Z_0^\ast \to Z_1 H^0$) fall with energy, while
those from vector boson fusion rise. Thus, the latter may have a better 
chance at CLIC. There are other channels with tiny contributions, like
$e^+e^-\to Z_2\to H^+H^-$, which we have not included in the analysis.
This channel, for example, is tiny even on the $Z_2$ resonance and 
completely negligible off it. 

\begin{figure}[htbp] 
\vspace{-10pt}
\centerline{\hspace{-3.3mm}
\rotatebox{-90}{\epsfxsize=6cm\epsfbox{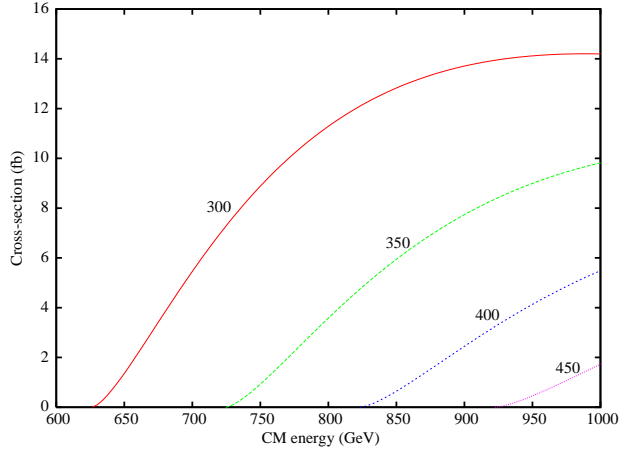}}}
\caption{Charged scalar pair production cross-section versus
the CM energy for different $1/R$. From top to bottom, the curves are
for $R^{-1}=$300, 350, 400, and 450 GeV respectively. $\Lambda R$ has been
fixed at 20, but the curves are not sensitive to the precise value of
the cutoff.}
\end{figure}

\begin{figure}[htbp] 
\vspace{-10pt}
\centerline{\hspace{-3.3mm}
\rotatebox{-90}{\epsfxsize=6cm\epsfbox{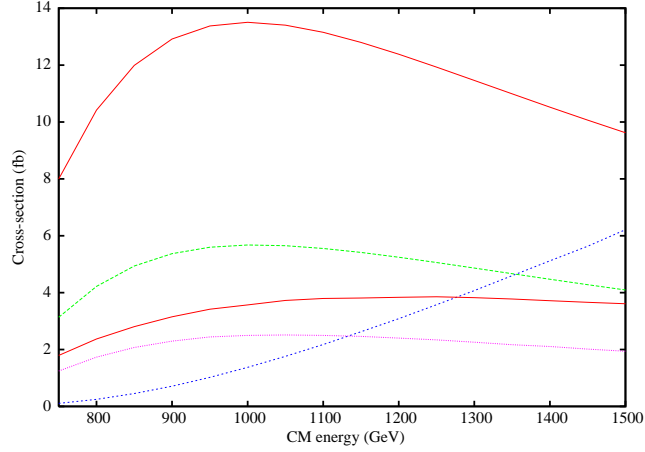}}}
\caption{Cross-section of various processes involving one or more excited
scalars. The plot is for $1/R=300$ GeV and $\Lambda R = 20$. From top to
bottom (at the right-hand edge) are the curves for $H^+H^-$, $H^0 \nu_0 
\bar{\nu_1}$ ($+$ h.c.), 
$H^0A^0$, $H^+ e_0 \bar{\nu_1}$, and $H^0 A^0 \gamma$.}
\end{figure}

\begin{figure}[htbp] 
\vspace{-10pt}
\centerline{\hspace{-3.3mm}
\rotatebox{-90}{\epsfxsize=6cm\epsfbox{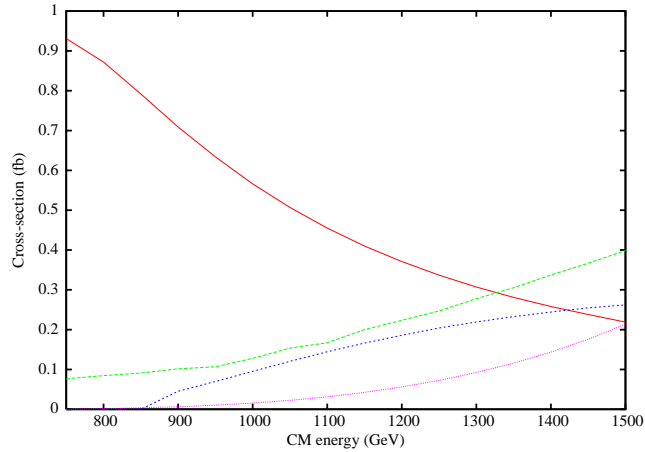}}}
\caption{Cross-section of various processes involving one or more excited
scalars. The plot is for $1/R=300$ GeV and $\Lambda R = 20$. From top to
bottom (at the right-hand edge) are the curves for $H^0 e_1^+ e_0^-$ ($+$
h.c.), $H^0 W_0^+
H^-$, $W_1^+ H^-$ (the falling one) and $Z_0 W_1^+ H^-$ (the rising one).}
\end{figure}

\begin{figure}[htbp] 
\vspace{-10pt}
\centerline{\hspace{-3.3mm}
\rotatebox{-90}{\epsfxsize=6cm\epsfbox{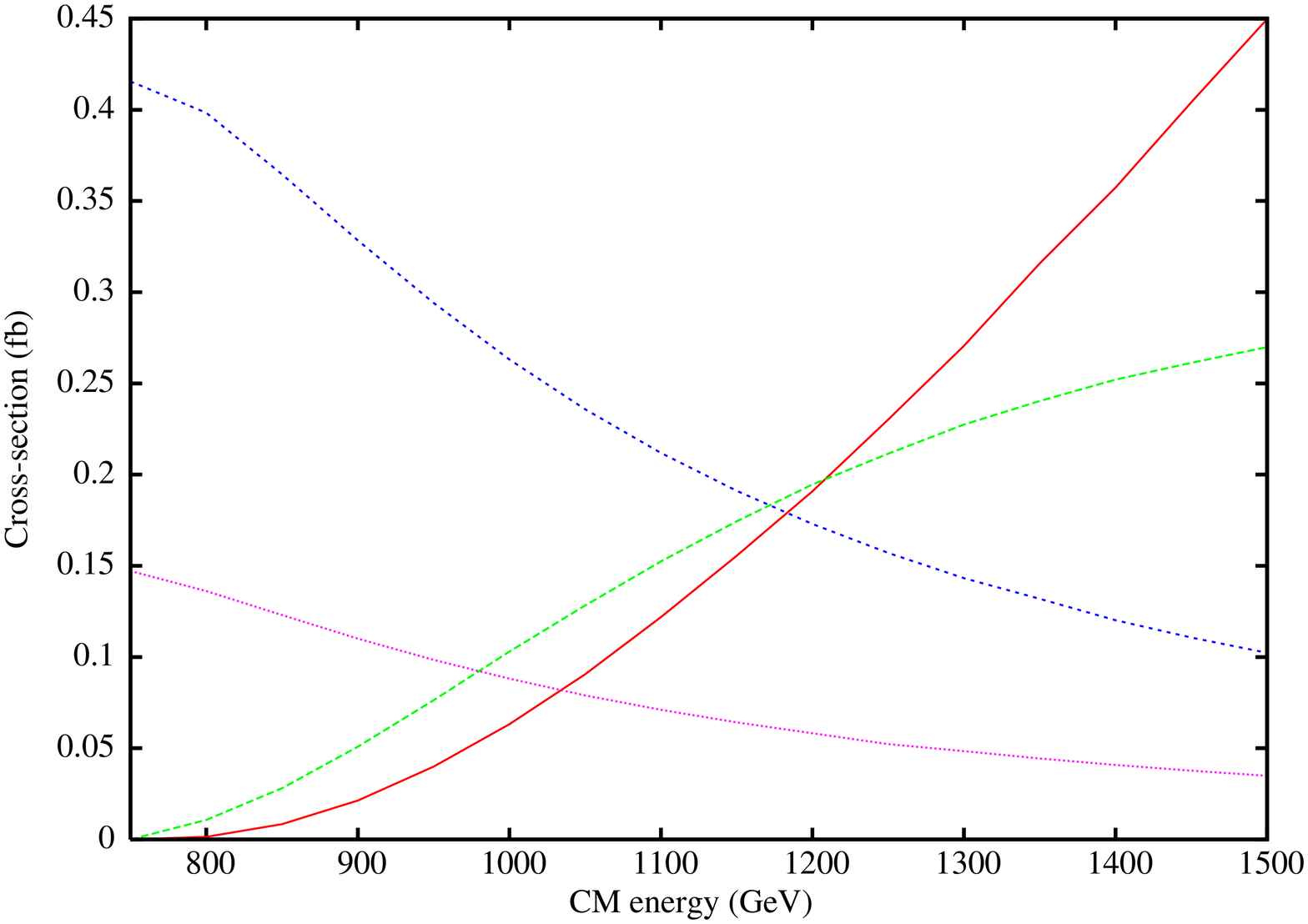}}}
\caption{Cross-section of various processes involving one or more excited
scalars. The plot is for $1/R=300$ GeV and $\Lambda R = 20$. From top to
bottom (at the right-hand edge) are the curves for $h W_1^+ H^-$, $W_0^+A^0 
H^-$, $Z_1H^0$, and $H^0 A^0$. $h$ is the SM Higgs boson.}
\end{figure}

We have also studied four-body processes like $e^+e^-\to H^+H^- \nu \bar\nu,
A^0A^0\nu \bar\nu, H^0H^0\nu \bar\nu$. These cross-sections are too small even
at ILC, and is without any hope of detection. CLIC may do a better
job, since these channels, mediated mostly by fusion of $n=1$ vector bosons,
rise with $\sqrt{s}$. The first channel may have a cross-section of 14 fb
at $\sqrt{s}=3$ TeV. 

\subsection{SM and UED Backgrounds}

The signal process is $e^+e^-\to n\tau + E\!\!\!/$, where $2\leq n \leq 4$.
Remember that these $\tau$s must be soft. For example, with $\sqrt{s}=1$ TeV,
$1/R=300$ GeV, $\Lambda R=20$, $\mhsqbar=0$, and $m_h=120$ GeV, the maximum
energy of the $n=0$ $\tau$s coming from $H^\pm$ decay is about 12 GeV. 
This value is sensitive to the excited scalar-$\tau$ mass splitting, which 
in turn depends on $\mhsqbar$.  

The SM background mostly comes from $W$-pair production, both of which decays
to $\tau$ leptons and $\tau$ neutrinos. These leptons are generally hard,
and this background can effectively be put under control if we apply an
upper energy cut of 12-15 GeV to the $\tau$ leptons. A subdominant background
comes from $Z$ pair production, where one of them decays to $\tau^+\tau^-$
and the other one decays invisibly. Such backgrounds may be eliminated by
reconstructing the $Z$. Similar considerations apply for 3-4 $\tau$ signals,
where a reconstruction can effectively eliminate such backgrounds.

Another prominent source of SM background is the $\gamma^*
\gamma^* \to \tau^+\tau^-$ events, where $\gamma^*$s originate from the
initial electron-positron pair which go undetected down the
beam pipe \cite{colorado}. The $\gamma^* \gamma^*$ production cross
section is $\sim 10^4$ pb. About half of these events results in final
state $e^+e^-$ pair as visible particles.
The background $\tau^+\tau^-$ pairs are usually quite soft and coplanar with
the beam axis. An acoplanarity cut significantly removes
this background. Such a cut, we have checked, does not appreciably
reduce our signal. For example, excluding events which deviate from
coplanarity within 40 mrad reduces only 7\% of the signal cross
section. In fact, current designs of ILC envisage very forward
detectors to specifically capture the `would-be-lost' $e^+e^-$ pairs
down the beam pipe

The UED backgrounds are more severe. There are at least three two-body processes
which can potentially swamp the signal: $e^+e^-\to W_1^+W_1^-, \tau_1^+\tau_1^-$
and $\tau_2^+\tau_2^-$. Among them the $W_1$ pair production cross-section
is largest \cite{rizzo}: at $1/R=300$ GeV, this is approximately 540 fb
for $\sqrt{s}=1$ TeV. Since $W_1$ must decay through leptonic channels, 
the cross-section for getting two $\tau$s plus missing energy in the final
state is about 60 fb. This can come from two different subprocesses: 
$W_1\to\tau_{n=1} \bar{\nu_\tau}_{n=0}$ 
and $W_1\to\tau_{n=0} \bar{\nu_\tau}_{n=1}$ (remember
that $n=1$ neutrinos are invisible). Whether one can apply a suitable 
upper energy cut depends on the precise position of these Higgses, {\em i.e.}, 
on $\mhsqbar$. For $\mhsqbar=0$ and $1/R=300$ GeV, the $\tau$s coming
from $W_1$ can have energies ranging from $2.8$ to 48 GeV. On the other
hand, $\tau$s coming from $H^+$ are softer, between $2.2$ and $3.3$ GeV.
Thus, one can avoid the $W_1$-background by putting an upper cut at $3.3$
GeV. This method fails completely if $\mhsqbar$ is large and positive,
say $10^4$ GeV$^2$; the reason is that $H^+$ becomes more massive, almost
degenerate with $W_1$. Unfortunately, we have no way to guess the value
of $\mhsqbar$ beforehand, so even the $W$-background removal is quite
difficult.

Similar backgrounds also come from $e^+e^-\to {\rm LKP}+Z_1$ and $e^+e^-
\to 2Z_1$. The first one is about five times smaller than the $W_1$ pair
production rate, but the decay nature of $Z_1$ (only to leptons) make the
$2\tau+E\!\!\!/$ background about one-fourth as significant as the other. 
$W_1$ and $Z_1$ being almost degenerate, the energies of the $\tau$s
are going to be similar. A much smaller background comes from $2Z_1$
channel. 

A more serious background comes from the excited $\tau$ ($\tau_1$ or $\tau_2$)
pair production through $s$-channel photon or $Z$ (unless one sits precisely
on the $\gamma_2$ or $Z_2$ resonances, their contribution can be neglected).
The $\tau_2$ pair production cross-section is about 100 fb for $1/R=300$ GeV,
and slightly less for $\tau_1$. (The difference is due to their couplings
with $Z$.)
It falls with increasing $1/R$, but so does the $H^\pm$ production rate.
The excited $\tau$s decay only to normal $\tau$s emitting an LKP, so the
background is identical to the signal, even in angular distribution. Again,
$\mhsqbar$ plays an important role: for $\mhsqbar=0$, $\tau_2$ backgrounds
can be removed by an identical mechanism to that of $W_1$. However, the
$\tau_1$ background is impossible to remove, or even to reduce significantly.
We show, in figures 5 and 6, the possible range of $\tau$ energies
coming from $W_1$, $\tau_1$, $\tau_2$, and $H^+$, for two values of
$\mhsqbar$: 0 and 5000 GeV$^2$.   

\begin{figure}[htbp] 
\vspace{-10pt}
\centerline{\hspace{-3.3mm}
\rotatebox{-90}{\epsfxsize=6cm\epsfbox{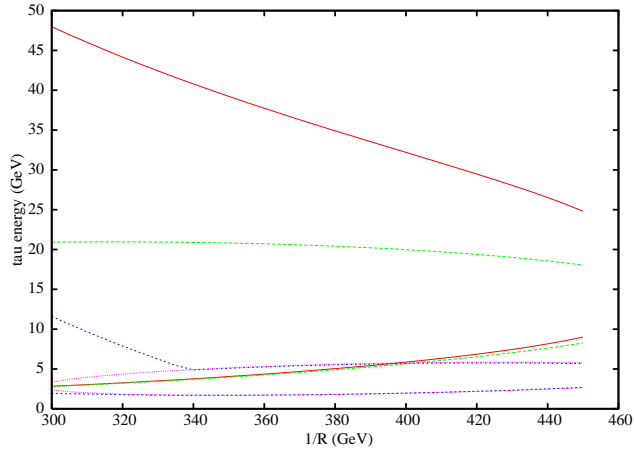}}}
\caption{The upper and lower energy bands of $n=0$ $\tau$ lepton coming 
from $W_1$ (solid black), $H^+$ (short-dashed blue), $\tau_2$ (long-dashed 
green) and $\tau_1$ (dotted magenta) decays. We have set $1/R=300$ GeV, 
$\Lambda R=20$, $\mhsqbar=0$ and $m_h=120$ GeV.}
   \label{fig5-0}
\end{figure}
\begin{figure}[htbp] 
\vspace{-10pt}
\centerline{\hspace{-3.3mm}
\rotatebox{-90}{\epsfxsize=6cm\epsfbox{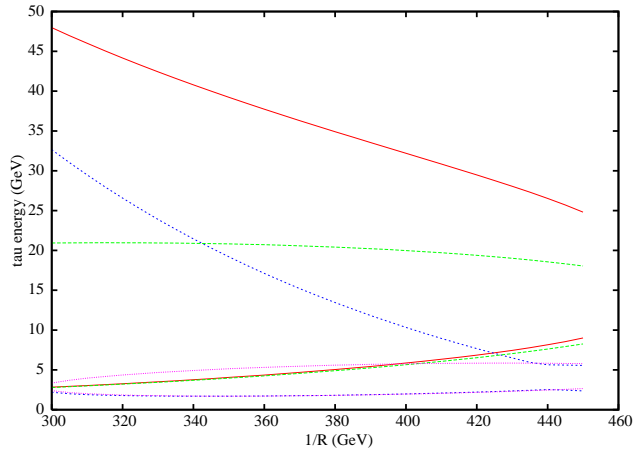}}}
\caption{Same as in fig.\ \ref{fig5-0}, but for $\mhsqbar=5000$ GeV$^2$.
All other legends remain the same.}
\end{figure}
This seems to be almost a no-go: if $W_1$ and $\tau_2$ backgrounds can
be removed, one gets stuck at $\tau_1$, or vice versa. For one of the most
favorable situations ($1/R=300$ GeV, $\Lambda R=20$, $\mhsqbar=0$, and
$m_h=120$ GeV), the signal, after providing optimum cuts, is at about $1\sigma$
level, assuming that we know $\mhsqbar$ beforehand, which is an
impossibility (unless there is some theoretical model). One way out may be
to use the muon channel, $e^+e^-\to \mu^+\mu^- + E\!\!\!/$, coming from
the production of excited muons only, as a calibration and look for
excess events in the $\tau$ channel.
 
There is also a case where the $\tau$s coming from $n=1$ Higgs 
decays cannot even
be detected. This happens when $m_{H^+}$ is close to $m_{\gamma_1}$. All
the excited scalars 
will lie close to the LKP, and the $\tau$s will be so soft as to
avoid detection. Of course, it may so happen that the $\tau$-window is
completely closed for $H^+$, which should then decay to $\mu^+$ with a 
much longer lifetime.

Thus, there are three levels of challenge. First, to have a theoretical
prediction for $\mhsqbar$, perhaps from some more fundamental theory. This
appears impossible at present, but we will try to get a bound on $\mhsqbar$
in the next section. Second is to detect the soft $\tau$s, which, hopefully,
is not a major problem; one can expect $\tau$s with energy more than $1.5$-2
GeV to be detected at the ILC. The third, which is the most challenging,
is the observation of the excited scalar sector. 

\section{Theoretical bound on $\mhsqbar$}

Since the radiative correction to the excited scalar masses is universal
in nature, it is evident from eq.\ (\ref{radcorhiggs}) that among those
scalars, $H^\pm$ will be the lowest-lying one. For sufficiently negative
values of $\mhsqbar$, the $H^+$ mass can go down below that of $\gamma_1$,
the LKP, and this sets the lower bound on $\mhsqbar$ as a function of
$1/R$ and $\Lambda$. 

\begin{figure}[htbp] 
\vspace{-10pt}
\centerline{\hspace{-3.3mm}
\rotatebox{-90}{\epsfxsize=6cm\epsfbox{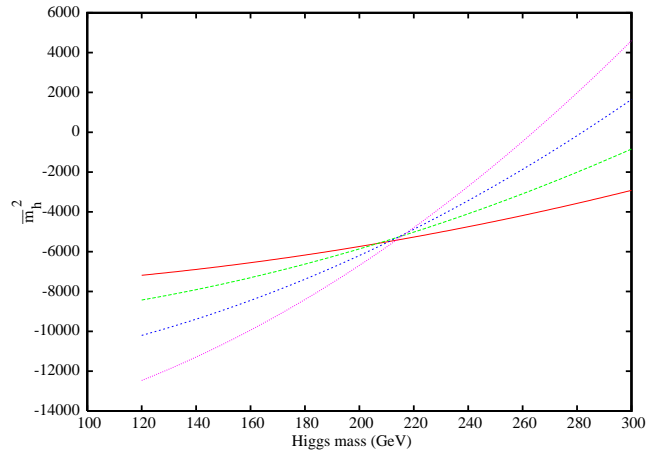}}}
\caption{The lower bound on $\mhsqbar$ as a function of the SM Higgs boson 
mass and $1/R$. $\Lambda R$ is fixed at 20. From top to
bottom (at the right-hand edge) are the curves for $1/R=$ 600, 500, 400, and
300 GeV respectively.}
\end{figure}
\begin{figure}[htbp] 
\vspace{-10pt}
\centerline{\hspace{-3.3mm}
\rotatebox{-90}{\epsfxsize=6cm\epsfbox{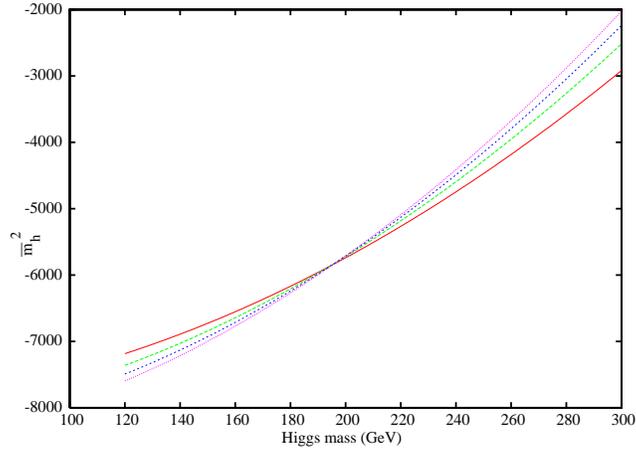}}}
\caption{The lower bound on $\mhsqbar$ as a function of the SM Higgs boson 
mass and $\Lambda R$, $1/R$ being fixed at 300 GeV. From top to
bottom (at the right-hand edge) are the curves for $\Lambda R=$ 50, 40,
30, and 20 respectively.}
\end{figure}

In fig.\ 7 we show the minimum allowed value of $\mhsqbar$ as a function of
$1/R$ and $m_h$. These curves are not overly sensitive to the precise
value of $\Lambda R$; the dependence is shown in fig.\ 8. Since $\mhsqbar$ 
cannot have a large negative value, for most of the parameter space the 
$\tau$ window will remain open. However, it can have large positive values.
For such values, the excited Higgs masses may go above the corresponding
electroweak gauge boson masses. This will open up a new set of decay channels,
like $H^0\to W_1^+ \ell \bar\nu_\ell$. They are comparable to the two-body
channels, since the coupling is a gauge one, not Yukawa. The final state
may not have any $\tau$-lepton in them. Since $W_1$ lies below the excited 
quarks, the final state must be hadronically quiet. Also note that for
large values of $1/R$ and a heavy SM Higgs boson, the lower limit on
$\mhsqbar$ may turn out to be positive. However, this region is mostly
outside the reach of ILC. 

\begin{figure}[htbp] 
\vspace{-10pt}
\centerline{\hspace{-3.3mm}
\rotatebox{-90}{\epsfxsize=6cm\epsfbox{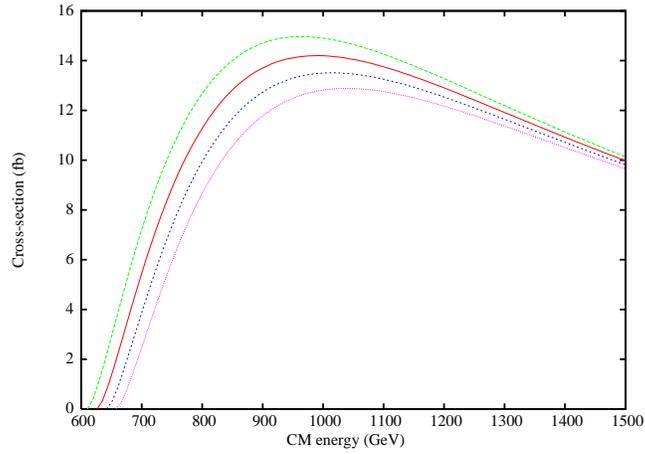}}}
\caption{The variation of cross-section for the process $e^+e^-
\to H^+H^-$ with $\mhsqbar$. From top to bottom, the curves are for
$\mhsqbar=-5000$, 0, 5000 and 10000 GeV$^2$ respectively. We have fixed
$1/R=300$ GeV, $\Lambda R=20$, and the SM Higgs boson mass $m_h=120$
GeV.}
\end{figure}

Is it possible to measure $\mhsqbar$ from precision studies at ILC? For that
matter, let us look at the most promising channel, pair production of $H^+
H^-$. In fig.\ 9 we show how the cross-section changes for different values
of $\mhsqbar$. The change is perceptible but may be too small for an
experimental detection (also note that the curves are drawn for a favourable
point in the parameter space, we may not be so lucky). The problem is further
aggravated by the fact that most of the energy in the final state is missing,
so it is almost impossible to reconstruct the $H^+$ invariant mass. Since 
leptons, quarks, and gauge bosons do not feel the effect of $\mhsqbar$, we
conclude that this is one parameter likely to remain unknown.
 
\section{Conclusion} 

In this paper we have focussed on the possible production, decay, and
detection for the $n=1$ excited scalar sector of the Universal Extra Dimension
model. These scalars will, with almost 100\% branching ratio, decay to soft  
$n=0$ $\tau$ leptons, accompanied by huge missing energy. Once a UED-type
new physics is established at the LHC and subsequently at the ILC (by the study
of KK leptons, quarks, and gauge bosons), it seems imperative to study the
scalar sector, at least to determine the third and last input parameter
to this model, namely, $\mhsqbar$, the boundary mass squared term for the
Higgs boson.

Unfortunately, such a thing is easier said than done. By itself, the soft
$\tau$ detection is a challenge, but most likely it will be overcome
at both LHC and ILC. The problem is to reduce the backgrounds. The SM
backgrounds can be removed by judiciously choosing the cuts, but the
backgrounds coming from other UED processes (pair production of $W_1$,
$\tau_1$, $\tau_2$, $Z_1\gamma_1$, etc.) are more difficult to handle.
For most of the parameter space, such processes give identical signals as
the excited scalar
production, and with almost identical energies and angular
distribution. Thus, it needs a precision study to detect those scalars,
and can be performed only at ILC, or CLIC. One may use the excited muon channel
as a calibration. The best channel to look for is the
charged scalar pair production, for which the signal may just be
detected.

Thus, $\mhsqbar$ seems to be one of the most challenging parameter to extract.
One can, of course, get a theoretical lower bound on $\mhsqbar$, stemming
from the fact that among all the scalars, $H^+$ is the lowest-lying, and for
sufficiently large negative values of $\mhsqbar$, it can go down below
$m_{\gamma_1}$ and become the LKP. This sets a lower bound on this term, which
is large and negative for most of the parameter space that can be probed at ILC.
However, this bound can even be positive for large values of $1/R$ and
a large SM Higgs mass. One can say more about this bound once the SM Higgs
boson is detected at the LHC.

\centerline{\bf Acknowledgements}

A.K.\ thanks the Department of
Science and Technology, Govt.\ of India, for the research project
SR/S2/HEP-15/2003. He also thanks the Theoretical Physics division of
Universit\"at Dortmund, where a part of the work was done,
for hospitality. B.B.\ thanks UGC, Govt.\ of India, for a research
fellowship.

\end{document}